\documentclass[conference]{IEEEtran}
\IEEEoverridecommandlockouts
\usepackage{cite}
\usepackage{amsmath,amssymb,amsfonts}
\usepackage{algorithmic}
\usepackage{graphicx}
\usepackage{textcomp}
\usepackage{xcolor}
\usepackage{hyperref}
\usepackage{multirow}
\def\BibTeX{{\rm B\kern-.05em{\sc i\kern-.025em b}\kern-.08em
    T\kern-.1667em\lower.7ex\hbox{E}\kern-.125emX}}
\begin{document}

\title{SymbolicPhasor: Power System Phasor Estimation via Deep Symbolic Regression\\
}

\author{\IEEEauthorblockN{Sina Mohammadi, Wencong Su}
\IEEEauthorblockA{\textit{Dept. of Electrical \& Computer Engineering} \\
\textit{University of Michigan-Dearborn}\\
Dearborn, USA \\
\{sinamo,wencong\}@umich.edu}

% \author{\IEEEauthorblockN{1\textsuperscript{st} Sina Mohammadi, Wencong Su}
% \IEEEauthorblockA{\textit{Dept. of Electrical \& Computer Engineering} \\
% \textit{University of Michigan-Dearborn}\\
% Dearborn, USA \\
% \{sinamo,wencong\}@umich.edu}
% \and
% \IEEEauthorblockN{2\textsuperscript{nd} Wencong Su}
% \IEEEauthorblockA{\textit{Dept. of Electrical \& Computer Engineering} \\
% \textit{University of Michigan-Dearborn}\\
% Dearborn, USA \\
% wencong@umich.edu}

% \and
% \and
% \and
% \and
}

\IEEEaftertitletext{\vspace{-1cm}}
\maketitle

\begin{abstract}
% Accurate phasor estimation is essential for reliable power system protection, particularly during fault conditions where signal distortions are prominent. However, its performance is degraded by the presence of decaying DC offsets (DCO), which arise due to transient phenomena in fault currents. Traditional techniques such as the Discrete Fourier Transform (DFT) are widely used for phasor estimation but are highly sensitive to DCO, resulting in significant magnitude and phase estimation errors, especially within the first cycle after a fault. To address these limitations, this paper proposes an enhanced Deep Symbolic Regression (DSR) framework for extracting the fundamental frequency component from distorted signals. By incorporating system frequency as symbolic constraints, the DSR model learns interpretable analytical expressions directly from fault current data. This enables effective suppression of DCO, harmonics, and noise within a single cycle, providing a precise, robust, and data-driven alternative for next-generation protection systems.

Accurate phasor estimation during power system faults is challenging because fault currents contain decaying DC offsets, harmonics, noise, and possible frequency deviations. These distortions can significantly degrade conventional discrete Fourier transform-based estimators, especially during the first cycle after fault inception. This paper presents SymbolicPhasor, a dynamic deep symbolic regression framework for estimating the fundamental component of distorted fault current signals. The method processes the signal through overlapping moving windows, learns interpretable analytical expressions for the full waveform within each window, and then projects the reconstructed signal onto nominal sine and cosine bases to recover the instantaneous fundamental magnitude and phase. By embedding symbolic tokens corresponding to the nominal, third-, and fifth-order harmonic frequencies, the proposed approach is guided toward physically meaningful expressions while preserving data-driven flexibility. The method is evaluated under single decaying-DC, multiple decaying-DC, and off-nominal frequency conditions. Results show consistently high reconstruction accuracy, with coefficient of determination values reaching 0.985, demonstrating that the proposed framework can recover the fundamental component within one cycle for practical protective relaying and measurement applications.
\end{abstract}


\vspace{-3mm}
\section{Introduction}
When a fault occurs in an electric power system, the resulting fault current deviates from a pure sinusoidal waveform due to the presence of a decaying DC offset (DCO). This component arises from the inherent inductive nature of power systems, where current cannot change instantaneously, and its magnitude is strongly influenced by the system’s X/R ratio. The DCO exhibits an aperiodic exponential behavior that cannot be directly captured by conventional phasor estimation techniques, thereby degrading estimation accuracy and potentially leading to malfunction or maloperation of protection and measurement devices \cite{al2011online,mohammadi2022novel}.

\begin{figure*}[t]
    \centering
    \includegraphics[width=0.75\linewidth]{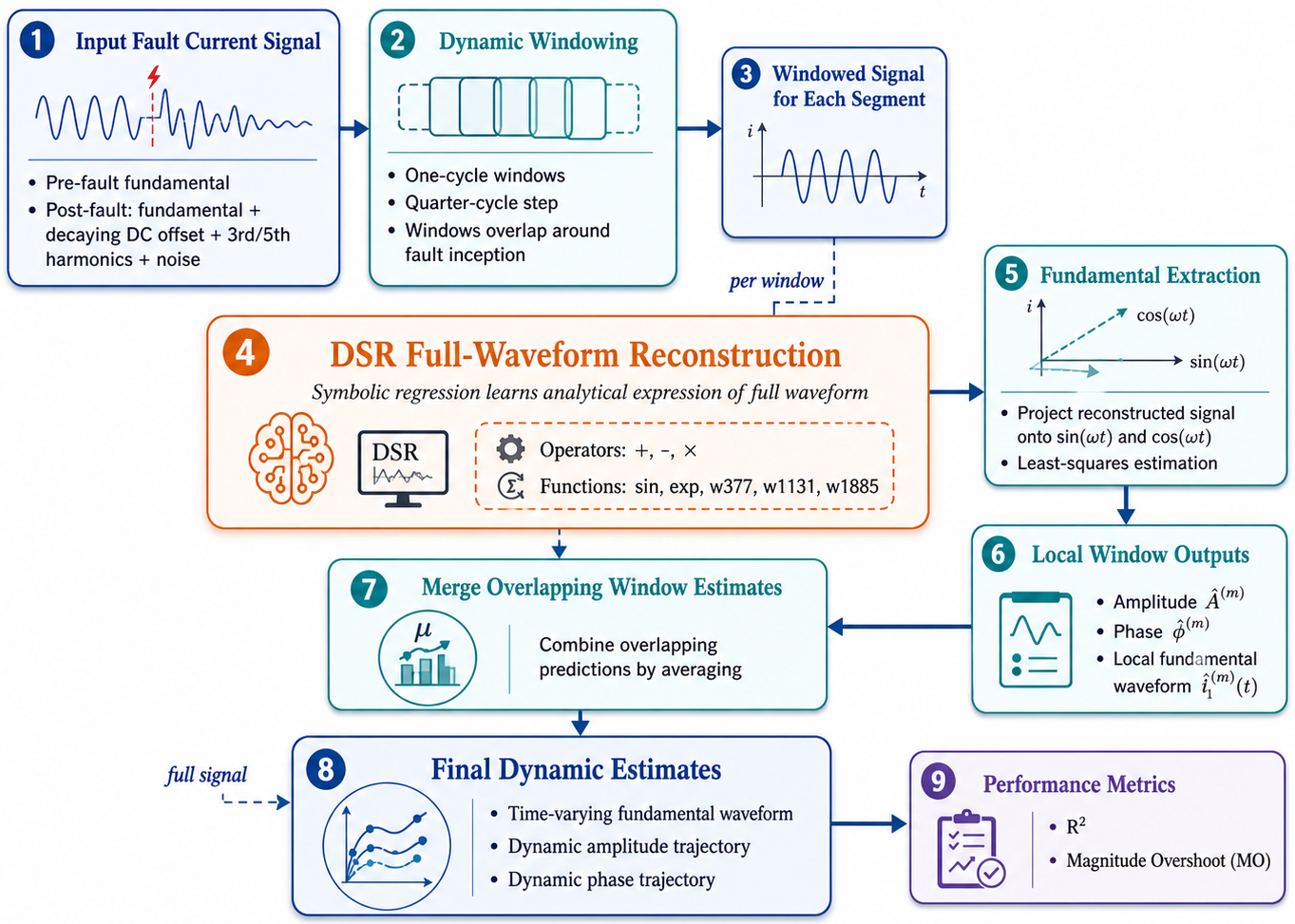}
    \caption{Overview of the proposed DSR-based framework for dynamic fundamental harmonic estimation of distorted fault current signal}
    \label{fig:proposed_method}
    \vspace{-3mm}
\end{figure*}

The discrete Fourier transform (DFT) is widely employed in protection applications for estimating fundamental and harmonic components due to its simplicity, robustness against higher-order harmonics, and fast computational response. Despite these advantages, DFT is highly sensitive to the presence of DCO, which it neither eliminates nor attenuates. As a result, phasor estimation errors can reach up to 15\%, leading to transient oscillations in magnitude estimation and degraded relay performance \cite{al2011online}. Therefore, effective mitigation of the DCO component prior to or during phasor estimation is essential to ensure accurate and reliable operation of protection systems.

A broad range of techniques has been proposed in the literature to address the DCO problem \cite{mohammadi2022decaying}. These methods can generally be classified into four main categories: pre-DFT, post-DFT, least-square (LS)-based, and artificial intelligence (AI)-based approaches. Each category offers distinct advantages but also suffers from inherent limitations in terms of accuracy, computational complexity, noise sensitivity, and data requirements.

In pre-DFT approaches, the DCO component is first estimated and removed from the fault current signal before applying the DFT. Several methods fall under this category, including Kalman-filter-based techniques, integral-based approaches, and Taylor series approximations \cite{abdoos2016accurate,jafarpisheh2016new,gu2000removal,yu2006discrete,girgis1982new}. These methods exploit properties such as the zero integral of sinusoidal components over one cycle to isolate the DCO component. However, they typically require at least one full cycle of data, which introduces delays in protection systems. Moreover, their accuracy deteriorates for small time constants, and they are highly sensitive to noise and modeling approximations \cite{eisa2008removal,abdoos2016accurate}. Additional techniques, such as dq0-based averaging, four-sample estimation, and down-sampling methods, also exhibit limitations including reduced accuracy under asymmetrical fault conditions and increased noise sensitivity due to limited sample usage \cite{soliman2004new,8352019,8610234}.

Post-DFT methods, in contrast, first compute the DFT and then estimate and remove the DCO from the obtained phasor quantities. Examples include mimic filters, cosine filters, and adaptive filtering techniques \cite{benmouyal2002removal,yu2006discrete}. While these methods can effectively compensate for DCO under certain conditions, their performance is often dependent on system parameters such as the time constant and X/R ratio. Additionally, many of these approaches require more than one cycle of data or rely on auxiliary signals such as voltage measurements, limiting their applicability in fast and practical protection schemes.

Least-squares (LS)-based methods estimate both harmonic components and DCO simultaneously by minimizing the error between measured signals and predefined mathematical models. These methods are attractive due to their relatively low computational burden and suitability for real-time implementation \cite{sachdev2007new}. However, their performance is often constrained by assumptions such as limited Taylor series expansions, predefined signal models, and sensitivity to noise and harmonic distortion \cite{jafarian2011weighted,hwang2018least,kim2019adaptive}. Advanced LS techniques, including iterative and adaptive variants, improve estimation accuracy but may introduce additional computational complexity and latency, making them less suitable for ultra-fast protection applications \cite{girgis1982new,sachdev2002recursive}.

In recent years, AI-based methods, including artificial neural networks and deep learning approaches, have gained attention for phasor estimation problems \cite{da2015phasor,kim2019study,sok2022deep,10591924}. These methods are capable of modeling nonlinear relationships and handling noise, harmonics, and complex signal characteristics effectively. However, they require extensive offline training, large datasets, and careful tuning of model parameters. Furthermore, many existing AI-based approaches do not fully account for practical factors such as multiple DCO components or variations in system frequency, which limits their generalization capability.

In this paper, we use deep symbolic regression (DSR) to estimate the fundamental harmonic of fault current signals. DSR is a machine learning approach that discovers mathematical expressions representing the underlying data using neural networks \cite{petersen2019deep,javadi2025review,10215513,10315150}. DSR is capable of extracting the fundamental component from current time-series signals by incorporating predefined symbolic tokens as constraints during training, specifically the angular fundamental frequency. In this way, the model learns to extract the sinusoidal fundamental component from distorted fault current signals.

\section{Proposed Method}
\label{sec:method}

This paper proposes a dynamic DSR framework for estimating the fundamental component of distorted fault current signals in power systems. Instead of relying on a single static fitting interval, the proposed method operates on overlapping time windows so that both pre-fault and post-fault behaviors are captured around the fault inception instant. Within each window, symbolic regression is used to reconstruct the complete current waveform, including the fundamental harmonic and DCO components. The reconstructed waveform is then projected onto the nominal fundamental basis to estimate the instantaneous amplitude and phase of the fundamental harmonic.

Fig.~\ref{fig:proposed_method} illustrates the overall workflow. First, the distorted current signal is divided into overlapping windows. Then, in each window, the DSR model learns an analytical expression for the full waveform rather than directly forcing a pure sinusoidal fit. Finally, the reconstructed signal is mapped onto the nominal fundamental basis to obtain the amplitude, phase angle, and fundamental waveform for that window. The local estimates from all windows are then merged to generate a dynamic time-varying estimate over the full signal duration.

\subsection{Fault Current Signal Model}

The fault current is modeled as a piecewise signal composed of a pre-fault fundamental component and a post-fault distorted component. Before the fault inception time $t_f$, the current is assumed to be dominated by the nominal fundamental component. After $t_f$, the signal includes a sudden change in the fundamental magnitude together with a decaying DC offset, odd harmonics, and measurement noise. Accordingly,
\begin{equation}
i(t)=
\begin{cases}
A_{1,\mathrm{pre}}\sin(\omega t+\phi), & t<t_f,\\
i_{\mathrm{post}}(t), & t\ge t_f,
\end{cases}
\label{eq:fault_model_piecewise}
\end{equation}
where
\begin{equation}
\begin{aligned}
i_{\mathrm{post}}(t)=\;&A_{1,\mathrm{post}}\sin(\omega t+\phi) \\
&+ A_{\mathrm{dc}} e^{-(t-t_f)/\tau_{\mathrm{dc}}} \\
&+ A_3 \sin(3\omega t+\phi_3) \\
&+ A_5 \sin(5\omega t+\phi_5) \\
&+ n(t).
\end{aligned}
\label{eq:postfault_model}
\end{equation}
Here, $A_{1,\mathrm{pre}}$ and $A_{1,\mathrm{post}}$ denote the pre-fault and post-fault fundamental amplitudes, respectively, $A_{\mathrm{dc}}$ and $\tau_{\mathrm{dc}}$ are the magnitude and time constant of the decaying DC offset, $A_3$ and $A_5$ are the amplitudes of the third- and fifth-order harmonics, $\omega = 2\pi f$ is the nominal angular frequency, $\phi$ is the phase angle of the fundamental component, and $n(t)$ represents white noise.

\subsection{Dynamic Windowing Strategy}

To track the signal evolution across the fault transition, the proposed method uses a moving-window mechanism. Let the complete sampled signal be denoted by $\{i[k]\}_{k=1}^{N}$ and let each local window contain $N_w$ samples with a step size of $N_s$ samples. The $m$th window is written as
\begin{equation}
\mathbf{i}^{(m)}=
\left[
i[k_m],\, i[k_m+1],\, \dots,\, i[k_m+N_w-1]
\right]^{\top},
\label{eq:window_signal}
\end{equation}
where consecutive windows overlap because $N_s < N_w$. In the implemented framework, the window length corresponds to one cycle and the step size corresponds to a quarter cycle, which ensures that some windows include both pre-fault and post-fault samples around $t_f$.

For each window, the method produces a local reconstructed waveform and a local fundamental estimate. Since neighboring windows overlap, multiple local predictions may exist for the same sample. These overlapping estimates are combined by
\begin{equation}
\hat{i}_{1}[k] =
\frac{
\sum\limits_{m \in \mathcal{W}(k)}
\hat{i}_{1}^{(m)}[k]
}{
\sum\limits_{m \in \mathcal{W}(k)} 1
},
\label{eq:window_average}
\end{equation}
where $\mathcal{W}(k)$ denotes the set of windows covering sample $k$.

\subsection{Symbolic Regression for Full-Signal Reconstruction}

Instead of directly forcing the estimator to output only the fundamental sinusoid, the proposed DSR stage first reconstructs the full waveform inside each window. Let the sampled time vector within a window be
\begin{equation}
\mathbf{t}^{(m)}=
[t_1^{(m)},\, t_2^{(m)},\, \dots,\, t_{N_w}^{(m)}]^{\top}.
\end{equation}
The symbolic regressor receives only time as input and seeks an analytical expression
\begin{equation}
\hat{i}^{(m)}(t)=f^{(m)}(t),
\label{eq:symbolic_full}
\end{equation}
where $f^{(m)}(\cdot)$ is built from a restricted operator library.

The binary operators are restricted to
\begin{equation}
\mathcal{B}=\{+,\,-,\,\times\},
\end{equation}
while the unary operators are defined as
\begin{equation}
\mathcal{U}=
\left\{
\sin(\cdot),\,
\exp(\cdot),\,
w_{377}(\cdot),\,
w_{1131}(\cdot),\,
w_{1885}(\cdot)
\right\}.
\label{eq:unary_set}
\end{equation}
The predefined symbolic tokens are
\begin{equation}
w_{377}(t)=377t,\quad
w_{1131}(t)=1131t,\quad
w_{1885}(t)=1885t.
\label{eq:tokens}
\end{equation}

These tokens correspond to the nominal fundamental, third harmonic, and fifth harmonic angular frequencies, respectively. Therefore, the search space is biased toward expressions that can represent the key components of the distorted current signal, namely the exponential decaying term and the dominant odd harmonics.

The resulting symbolic model in each window can therefore approximate
\begin{equation}
\begin{aligned}
\hat{i}^{(m)}(t) \approx\;&
\alpha_1 \sin(377t+\beta_1) \\
&+ \alpha_3 \sin(1131t+\beta_3) \\
&+ \alpha_5 \sin(1885t+\beta_5) \\
&+ \alpha_{\mathrm{dc}} \exp(\beta_{\mathrm{dc}} t),
\end{aligned}
\label{eq:approx_form}
\end{equation}
although the exact discovered expression is determined automatically by the symbolic regression optimizer.

\subsection{Fundamental Component Extraction from the Reconstructed Signal}

After the DSR model reconstructs the windowed waveform, the fundamental component is extracted from the reconstructed signal using orthogonal projection onto the nominal fundamental basis. For the $m$th window, define
\begin{equation}
s^{(m)}(t)=\sin(\omega t), \qquad
c^{(m)}(t)=\cos(\omega t),
\end{equation}
where $\omega = 2\pi f$. The reconstructed signal $\hat{i}^{(m)}(t)$ is approximated as
\begin{equation}
\hat{i}^{(m)}(t)
\approx
a_{\sin}^{(m)} \sin(\omega t)
+
b_{\cos}^{(m)} \cos(\omega t).
\label{eq:ls_model}
\end{equation}

The coefficients are obtained by least squares:
\begin{equation}
\begin{bmatrix}
a_{\sin}^{(m)}\\
b_{\cos}^{(m)}
\end{bmatrix}
=
\arg\min_{a,b}
\left\|
\hat{\mathbf{i}}^{(m)}
-
a\,\mathbf{s}^{(m)}
-
b\,\mathbf{c}^{(m)}
\right\|_2^2.
\label{eq:ls_estimation}
\end{equation}

The estimated amplitude and phase are then
\begin{equation}
\hat{A}^{(m)}
=
\sqrt{
\left(a_{\sin}^{(m)}\right)^2
+
\left(b_{\cos}^{(m)}\right)^2
},
\label{eq:amp_est}
\end{equation}
\begin{equation}
\hat{\phi}^{(m)}
=
\tan^{-1}
\left(
\frac{b_{\cos}^{(m)}}{a_{\sin}^{(m)}}
\right).
\label{eq:phase_est}
\end{equation}
Finally, the local reconstructed fundamental waveform is
\begin{equation}
\hat{i}_1^{(m)}(t)=
\hat{A}^{(m)}
\sin\!\left(\omega t+\hat{\phi}^{(m)}\right).
\label{eq:fundamental_local}
\end{equation}

% \subsection{Reference DFT-Based Dynamic Estimator}

% To provide a fair comparison, a reference dynamic estimator based on the same moving-window structure is also considered. In each window, the measured signal is directly projected onto the nominal sine and cosine bases at $\omega$, and the amplitude and phase are computed using \eqref{eq:ls_estimation}--\eqref{eq:phase_est}. Therefore, the main difference between the two methods is that the DFT-based baseline applies the projection directly to the measured signal, whereas the proposed method first reconstructs the signal through DSR and then extracts the fundamental from the reconstructed waveform.

% \subsection{Performance Metrics}

% Let $i_1(t)$ be the true fundamental signal and $\hat{i}_1(t)$ be its estimate. The mean squared error (MSE) and root mean squared error (RMSE) are defined as
% \begin{equation}
% \mathrm{MSE}=
% \frac{1}{N}
% \sum_{k=1}^{N}
% \left(i_1[k]-\hat{i}_1[k]\right)^2,
% \end{equation}
% \begin{equation}
% \mathrm{RMSE}=\sqrt{\mathrm{MSE}}.
% \end{equation}
% The mean amplitude error is computed as
% \begin{equation}
% e_A=
% \frac{1}{N}
% \sum_{k=1}^{N}
% \left|\hat{A}[k]-A[k]\right|,
% \end{equation}
% and the mean wrapped phase error is given by
% \begin{equation}
% e_{\phi}=
% \frac{1}{N}
% \sum_{k=1}^{N}
% \left|
% \angle\!\left(
% e^{j(\hat{\phi}[k]-\phi[k])}
% \right)
% \right|.
% \end{equation}
% In addition, the coefficient of determination $R^2$, the total vector error (TVE), and the magnitude overshoot (MO) are used to quantify the overall fitting quality and phasor estimation accuracy.

\subsection{Performance Metrics}

Let $i_1(t)$ be the true fundamental signal and $\hat{i}_1(t)$ be its estimate. To evaluate the estimation accuracy, the coefficient of determination $R^2$ is used, which is defined as
\begin{equation}
R^2=
1-
\frac{\sum_{k=1}^{N}\left(i_1[k]-\hat{i}_1[k]\right)^2}
{\sum_{k=1}^{N}\left(i_1[k]-\bar{i}_1\right)^2},
\end{equation}
where $\bar{i}_1=\frac{1}{N}\sum_{k=1}^{N} i_1[k]$ is the mean of the true fundamental signal.

In addition, the magnitude overshoot (MO) is used to quantify the transient overestimation in the estimated amplitude trajectory. It is defined as
\begin{equation}
\mathrm{MO}=
\frac{\max\!\left(\hat{A}[k]\right)-A_{\mathrm{ref}}}
{A_{\mathrm{ref}}}\times 100\%,
\end{equation}
where $\hat{A}[k]$ is the estimated amplitude at sample $k$, and $A_{\mathrm{ref}}$ is the reference fundamental amplitude. These two metrics jointly assess the overall signal fitting quality and the transient amplitude behavior of the proposed method.

% \vspace{-5mm}
\section{Numerical Experiments}

This section evaluates the performance of the proposed DSR framework for fundamental phasor estimation under multiple fault conditions. The experiments are designed to assess robustness against (i) single DCO, (ii) multiple DCO components, and (iii) off-nominal system frequencies. All simulations are consistent with the signal model introduced in Section II and are implemented using a symbolic regression engine based on \texttt{PySR} library in \texttt{Python}.

% \vspace{-3mm}
% \subsection{Simulation Setup}

% The fault current signal is generated according to \eqref{eq:fault_model_piecewise} and \eqref{eq:postfault_model}, including fundamental, harmonic, DCO, and noise components. A nominal system frequency of $f=60$ Hz is assumed, corresponding to an angular frequency $\omega \approx 377$ rad/s. Each signal consists of one cycle sampled at 200 samples per cycle (12 kHz sampling rate), ensuring high temporal resolution for sub-cycle estimation.

% The symbolic regression model receives only the time variable $t$ as input and is constrained using a predefined operator set consisting of $\{+, -, \times, \exp(\cdot), \sin(\cdot), w_{377}(\cdot), w_{1131}(\cdot), w_{1885}(\cdot)$, where $w_{377}(t)=377t$ enforces the nominal frequency and $w_{1131}(t)=1131t, w_{1885}(t)=1885t$ enforces the higher order harmonics. The regression is trained for 25 iterations with a population size of 25 and a maximum expression size of 20.

% After symbolic regression, the extracted signal is projected onto orthogonal sinusoidal bases to obtain amplitude and phase estimates.

% The considered fault current scenarios are summarized in Table~\ref{tab:signal_params}. 
% These scenarios are designed to consider a range of possible fault current scenarios, including single and multiple DCO components as well as off-nominal frequency conditions.

\subsection{Simulation Setup}

The fault current signal is generated according to the piecewise model in (\ref{eq:fault_model_piecewise})--(\ref{eq:postfault_model}), including a pre-fault fundamental component and a post-fault distorted component composed of the fundamental harmonic, DCO, odd harmonics, and additive noise. A nominal system frequency of $f=60$ Hz is assumed, corresponding to an angular frequency of $\omega \approx 377$ rad/s. To evaluate the proposed dynamic framework around the fault transition, each simulated waveform contains both pre-fault and post-fault intervals, rather than only a single-cycle segment. The signal is sampled at 200 samples per cycle, which corresponds to a sampling frequency of 12 kHz.

Following Section II-B, the proposed method operates on overlapping moving windows. In the implemented framework, each local window has a length of one cycle, while the step size between consecutive windows is one quarter cycle. Therefore, neighboring windows overlap and some of them naturally contain both pre-fault and post-fault samples around the fault inception instant. For each window, the symbolic regression model reconstructs the full local waveform, after which the reconstructed signal is projected onto the nominal sine and cosine bases to estimate the local fundamental amplitude and phase. The overlapping local estimates are then merged to obtain a dynamic time-varying estimate over the full signal duration.

The symbolic regression model receives only the time variable $t$ as input and is constrained using the predefined operator set
\[
\{+, -, \times, \exp(\cdot), \sin(\cdot), w377(\cdot), w1131(\cdot), w1885(\cdot)\},
\]
where $w377(t)=377t$ enforces the nominal fundamental frequency, while $w1131(t)=1131t$ and $w1885(t)=1885t$ guide the symbolic search toward the third- and fifth-order harmonic components, respectively. The regression is trained for 25 iterations with a population size of 25 and a maximum expression size of 20.

After symbolic regression, the reconstructed waveform in each window is projected onto orthogonal sinusoidal bases at the nominal frequency to obtain the corresponding amplitude and phase estimates of the fundamental component.

The considered fault current scenarios are summarized in Table~\ref{tab:signal_params}. These scenarios are designed to cover representative fault conditions, including single-DCO, multiple-DCO, and off-nominal frequency cases.

For all simulation scenarios, the DFT method is used as the baseline for phasor computation comparison. 

% \begin{table*}[t]
% \centering
% \scriptsize
% \caption{Parameters of considered fault current scenarios}
% \label{tab:signal_params}
% \begin{tabular}{c c c c c c c c}
% \hline
% Case & Scenario & Frequency (Hz) & No. of DCOs & DCO Magnitudes (p.u.) & Time Constants (s) & Harmonics & Gaussian Noise \\
% \hline

% \multirow{3}{*}{Case 1 (Single DCO)} 
%  & S1 & 60 & 1 & 0.8 & 0.01 & 3rd, 5th &  $\sigma=0.01$ p.u. \\
%  & S2 & 60 & 1 & 0.4 & 0.02 & 3rd, 5th & $\sigma=0.01$ p.u. \\
%  & S3 & 60 & 1 & 0.6 & 0.04 & 3rd, 5th &  $\sigma=0.01$ p.u. \\

% \hline

% \multirow{3}{*}{Case 2 (Multiple DCO)} 
%  & S1 & 60 & 2 & 0.8, 0.35 & 0.01, 0.1 & 3rd, 5th & $\sigma=0.01$ p.u. \\
%  & S2 & 60 & 2 & 0.4, 0.20 & 0.02, 0.2 & 3rd, 5th &  $\sigma=0.01$ p.u. \\
%  & S3 & 60 & 2 & 0.6, 0.25 & 0.04, 0.4 & 3rd, 5th &  $\sigma=0.01$ p.u. \\

% \hline

% \multirow{2}{*}{Case 3 (Off-nominal frequency)} 
%  & S1 & 59.5 & 1 & 0.4 & 0.02 & 3rd, 5th & $\sigma=0.01$ p.u. \\
%  & S2 & 60.5 & 1 & 0.6 & 0.04 & 3rd, 5th & $\sigma=0.01$ p.u. \\

% \hline
% \end{tabular}
% % \vspace{-5mm}
% \end{table*}

\begin{table*}[t]
\centering
\scriptsize
\caption{Parameters of considered fault current scenarios}
\label{tab:signal_params}
\begin{tabular}{c c c c c c c c}
\hline
Case & Scenario & Frequency (Hz) & No. of DCOs & DCO Magnitudes (p.u.) & Time Constants (ms) & Harmonics (p.u.) & Gaussian Noise \\
\hline

\multirow{3}{*}{Case 1 (Single DCO)} 
 & S1 & 60   & 1 & 5           & 10        & 3rd $(0.5)$, 5th $(0.2)$  & SNR $=30$ dB \\
 & S2 & 60   & 1 & 5           & 20        & 3rd $(0.5)$, 5th $(0.2)$  & SNR $=30$ dB \\
 & S3 & 60   & 1 & 5           & 40        & 3rd $(0.5)$, 5th $(0.2)$  & SNR $=30$ dB \\

\hline

\multirow{3}{*}{Case 2 (Multiple DCO)} 
 & S1 & 60   & 2 & 5, 1        & 10, 100  & 3rd $(0.5)$, 5th $(0.2)$  & SNR $=30$ dB \\
 & S2 & 60   & 2 & 3, 0.5      & 20, 200  & 3rd $(0.5)$, 5th $(0.2)$  & SNR $=30$ dB \\
 & S3 & 60   & 2 & 2, 0.25     & 40, 400  & 3rd $(0.5)$, 5th $(0.2)$  & SNR $=30$ dB \\

\hline

\multirow{2}{*}{Case 3 (Off-nominal frequency)} 
 & S1 & 59.5 & 1 & 5           & 50        & 3rd $(0.5)$, 5th $(0.2)$  & SNR $=30$ dB \\
 & S2 & 60.5 & 1 & 5           & 50        & 3rd $(0.5)$, 5th $(0.2)$  & SNR $=30$ dB \\

\hline
\end{tabular}
\end{table*}

\vspace{-3mm}
\subsection{Case Study 1: Single DCO with Different Time Constants}

In the first experiment, the fault current includes a single decaying DC component, along with third- and fifth-order harmonics and additive Gaussian noise. Three representative scenarios are considered with varying time constants and phase angles.

\begin{figure}[t]
    \centering
    \includegraphics[width=0.95\columnwidth]{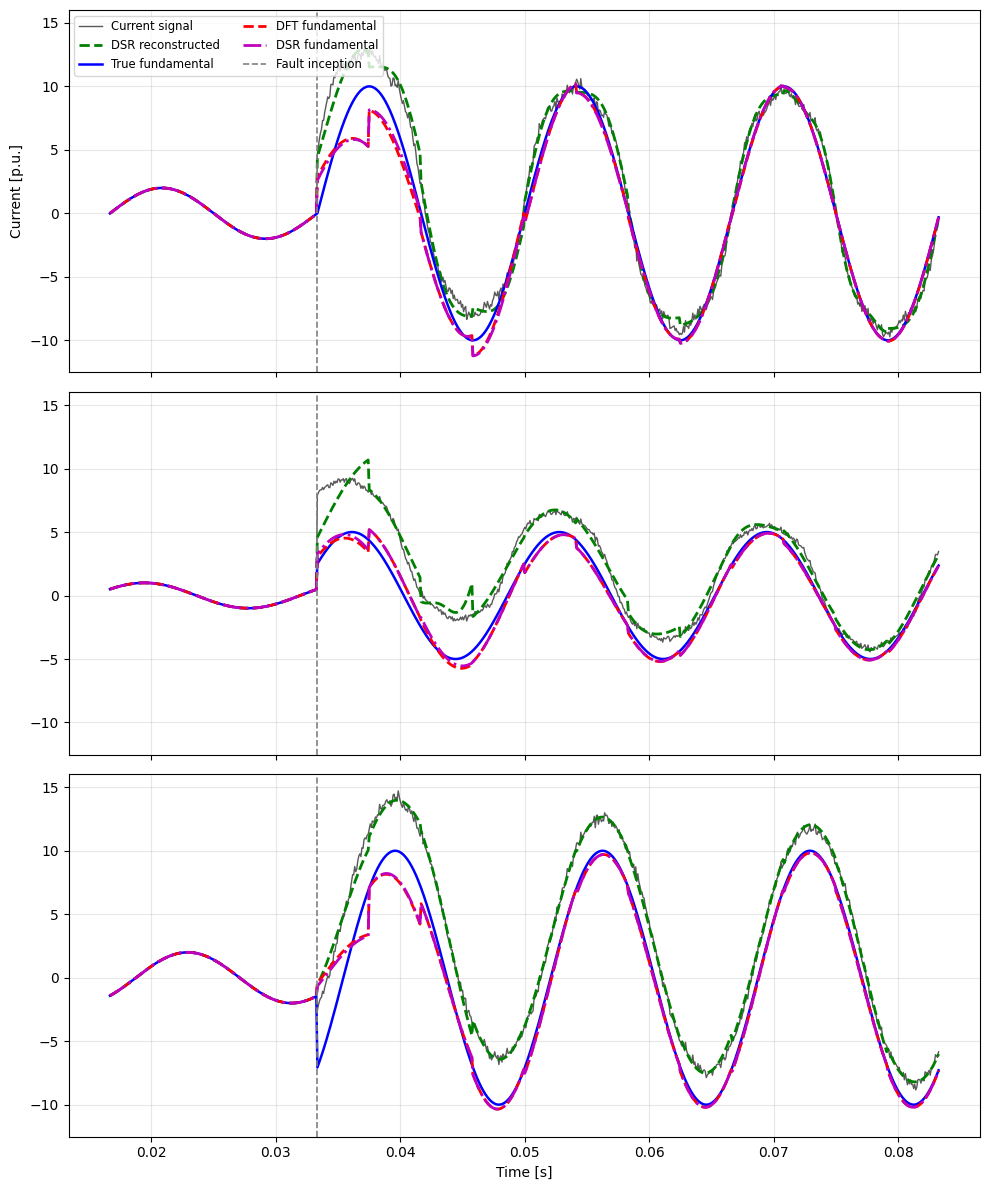}
    \caption{Fundamental frequency extraction under single DCO component. Scenarios S1, S2, and S3 are shown from top to bottom.}
    \label{fig:single_dco_results}
    \vspace{-5mm}
\end{figure}

Fig.~\ref{fig:single_dco_results} illustrates the distorted fault current signals and the reconstructed fundamental components obtained using the proposed DSR method. Despite the presence of DCO and harmonic distortions, the model extracts the fundamental component similar to DFT approach.

The quantitative results, summarized in Table~\ref{tab:single_dco_metrics}, show near-perfect reconstruction performance with high $R^2$ values and low MO across all scenarios.

\begin{table}[ht]
\centering
\caption{Performance under single DCO scenarios}
\label{tab:single_dco_metrics}
\begin{tabular}{c c c}
\hline
Scenario & $R^2$ & MO (\%)\\
\hline
S1 & 0.945 & 12.26 \\
S2 & 0.955 & 17.02 \\
S3 & 0.945 & 3.63 \\
\hline
\end{tabular}
% \vspace{-3mm}
\end{table}

\subsection{Case Study 2: Multiple DCO Components}

The second experiment introduces two decaying DC components with different time constants, representing the effect of current transformer saturation. The second DCO has smaller amplitude and larger time constant compared to the first one \cite{10591924}.

\begin{figure}[ht]
    \centering
    \includegraphics[width=0.95\columnwidth]{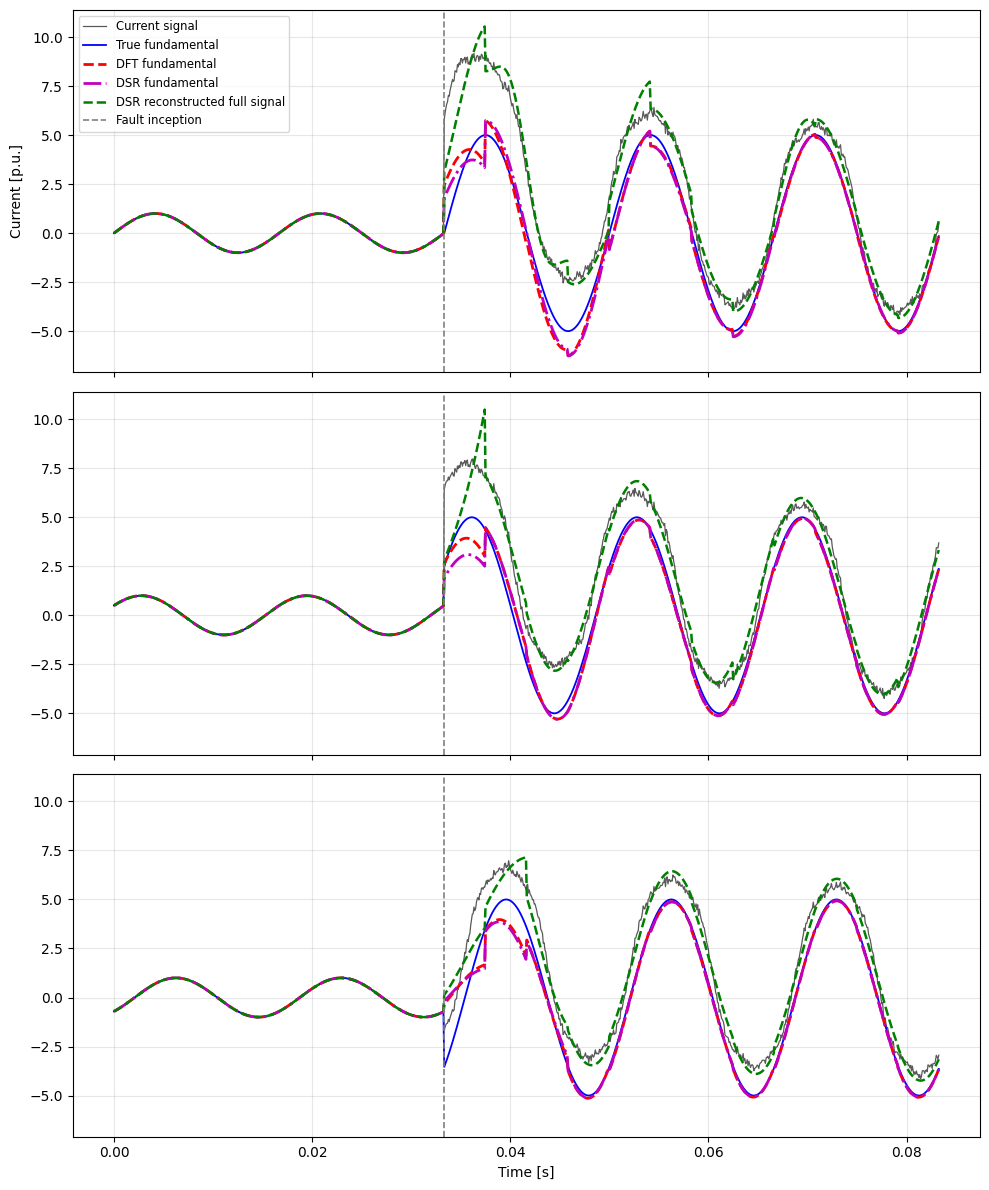}
    \caption{Fundamental frequency extraction across multiple DCO components. Scenarios S1, S2, and S3 are shown from top to bottom.}
    \label{fig:multi_dco_results}
    % \vspace{-3mm}
\end{figure}

Fig.~\ref{fig:multi_dco_results} shows the reconstruction results under these conditions. The presence of multiple exponential components significantly increases signal complexity, making traditional methods more susceptible to estimation errors.

Table~\ref{tab:multi_dco_metrics} reports the corresponding performance metrics. These results highlight the capability of DSR to handle multiple transient components simultaneously.

\begin{table}[ht]
\centering
\caption{Performance under multiple DCO scenarios}
\label{tab:multi_dco_metrics}
\begin{tabular}{c c c c}
\hline
Scenario & $R^2$ & MO (\%)\\
\hline
S1 & 0.979 & 22.97  \\
S2 & 0.977 & 8.87 \\
S3 & 0.966 & 8.92\\
\hline
\end{tabular}
\vspace{-3mm}
\end{table}

\subsection{Case Study 3: Off-Nominal Frequency Conditions}

In power systems, the operating frequency may deviate from its nominal value due to disturbances \cite{rogers2000power}. To evaluate the robustness of the proposed method, the third experiment considers off-nominal frequencies of 59.5 Hz and 60.5 Hz. Importantly, the symbolic regression model remains constrained to the nominal operator $w_{377}(t)$, thereby testing its ability to generalize under frequency mismatch.

\begin{figure}[ht]
    \centering
    \includegraphics[width=0.95\columnwidth]{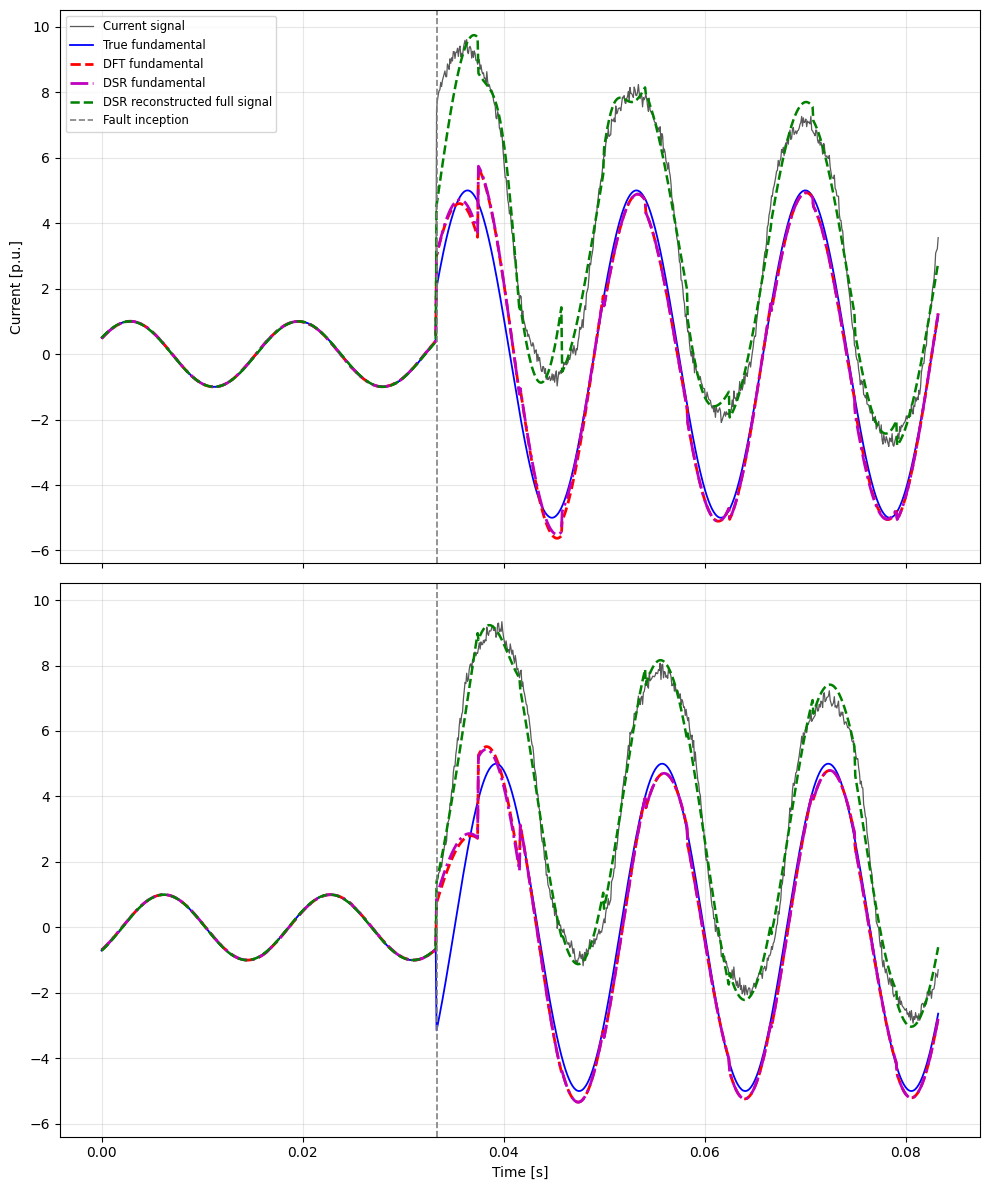}
    \caption{Fundamental extraction under off-nominal frequency conditions.}
    \label{fig:off_nominal_results}
    % \vspace{-5mm}
\end{figure}

Fig.~\ref{fig:off_nominal_results} presents the extraction results under these conditions. While the reconstructed signals remain close to the true fundamental component, slight deviations are observed due to the imposed frequency constraint.

% \begin{figure}[t]
% \centering
% \begin{tabular}{c}
%       \includegraphics[width=\linewidth]{off_nominal_59.5.png} \\
%       (a) \\
%       \includegraphics[width=\linewidth]{off_nominal_60.5.png}\\
%       (b) \\
% \end{tabular}
% \caption{Fundamental extraction under off-nominal frequency conditions. (a) 59.5 Hz (b) 60.5 Hz}
%     \label{fig2}
%     \vspace{-6mm}
% \end{figure}

Table~\ref{tab:off_nominal_metrics} summarizes the performance across scenarios. The results indicate that although the model is constrained to the nominal frequency, it still provides accurate approximations of the fundamental component. This suggests that the proposed approach is robust to moderate frequency deviations, which are common in real-world systems.

\begin{table}[t]
\centering
\caption{Performance under off-nominal frequency scenarios}
\label{tab:off_nominal_metrics}
\begin{tabular}{c c c c}
\hline
Scenario & Frequency (Hz)  & $R^2$ & MO (\%)\\
\hline
S1 & 59.5  & 0.985 & 14.33 \\
S2 & 60.5  & 0.951 & 8.99\\
\hline
\end{tabular}
\vspace{-3mm}
\end{table}

\section{Conclusion}

This paper presented SymbolicPhasor, a dynamic DSR framework for estimating the fundamental component of distorted fault current signals. The proposed method reconstructs the full waveform in overlapping moving windows and then extracts the nominal fundamental magnitude and phase through projection onto sinusoidal bases. By incorporating symbolic tokens associated with the fundamental, third-, and fifth-order harmonic frequencies, the method is guided toward physically meaningful expressions that capture the main components of fault current signals.

The numerical results showed that the proposed framework can accurately recover the fundamental component under single-DCO, multiple-DCO, and off-nominal frequency conditions. High $R^2$ values and low magnitude overshoot across the considered scenarios demonstrate the effectiveness of the method under severe waveform distortion.

The main limitation of the proposed approach is the computational burden of symbolic regression in each window. Future work will focus on improving computational efficiency and extending the framework to more realistic fault conditions.

% \vspace{-3mm}
\section*{Acknowledgments}
This work was supported by the University of Michigan-Dearborn Experience+ Student Independent Research Grant. The authors also acknowledge the use of ChatGPT for English writing assistance and grammar checking in the preparation of this manuscript.

\vspace{-3mm}
\bibliographystyle{IEEEtran}
\bibliography{Ref}

@article{al2011online,
  title={Online algorithm for removal of decaying DC-offset from fault currents},
  author={Al-Tallaq, Kamel NA and Al-Sharai, HD and El-Hawary, ME},
  journal={Electric power systems research},
  volume={81},
  number={7},
  pages={1627--1629},
  year={2011},
  publisher={Elsevier}
}

@article{abdoos2016accurate,
  title={Accurate and fast DC offset removal method for digital relaying schemes},
  author={Abdoos, Ali Akbar and Gholamian, Seyyed Asghar and Farzinfar, Mehdi},
  journal={IET Generation, Transmission \& Distribution},
  volume={10},
  number={8},
  pages={1769--1777},
  year={2016},
  publisher={Wiley Online Library}
}

@article{jafarpisheh2016new,
  title={A new DFT-based phasor estimation algorithm using high-frequency modulation},
  author={Jafarpisheh, Babak and Madani, Seyed M and Shahrtash, S Mohammad},
  journal={IEEE Transactions on Power Delivery},
  volume={32},
  number={6},
  pages={2416--2423},
  year={2016},
  publisher={IEEE}
}

@article{gu2000removal,
  title={Removal of DC offset in current and voltage signals using a novel Fourier filter algorithm},
  author={Gu, Jyh-Cherng and Yu, Sun-Li},
  journal={IEEE Transactions on Power Delivery},
  volume={15},
  number={1},
  pages={73--79},
  year={2000},
  publisher={IEEE}
}

@article{yu2006discrete,
  title={A discrete Fourier transform-based adaptive mimic phasor estimator for distance relaying applications},
  author={Yu, Chi-Shan},
  journal={IEEE Transactions on Power delivery},
  volume={21},
  number={4},
  pages={1836--1846},
  year={2006},
  publisher={IEEE}
}

@article{girgis1982new,
  title={A new Kalman filtering based digital distance relay},
  author={Girgis, Adly A},
  journal={IEEE Transactions on Power Apparatus and Systems},
  number={9},
  pages={3471--3480},
  year={1982},
  publisher={IEEE}
}

@inproceedings{eisa2008removal,
  title={Removal of decaying DC offset in current signals for power system phasor estimation},
  author={Eisa, Amir AA and Ramar, K},
  booktitle={2008 43rd International Universities Power Engineering Conference},
  pages={1--4},
  year={2008},
  organization={IEEE}
}

@article{soliman2004new,
  title={A new digital transformation for harmonics and DC offset removal for the distance fault locator algorithm},
  author={Soliman, SA and Alammari, RA and El-Hawary, ME},
  journal={International journal of electrical power \& energy systems},
  volume={26},
  number={5},
  pages={389--395},
  year={2004},
  publisher={Elsevier}
}

@ARTICLE{8610234,
  author={Kumar, Bandi Ravi and Kumar, Avinash},
  journal={IEEE Transactions on Power Delivery}, 
  title={Mitigation of the DC Offset by a Sub-Cycle Sample Method M-Class PMUs}, 
  year={2019},
  volume={34},
  number={2},
  pages={780-783},
  doi={10.1109/TPWRD.2019.2892600}}

@ARTICLE{8352019,
  author={Jafarpisheh, Babak and Madani, Seyed M. and Jafarpisheh, Siamak},
  journal={IEEE Transactions on Power Delivery}, 
  title={Improved DFT-Based Phasor Estimation Algorithm Using Down-Sampling}, 
  year={2018},
  volume={33},
  number={6},
  pages={3242-3245},
  doi={10.1109/TPWRD.2018.2831005}}

@article{benmouyal2002removal,
  title={Removal of DC-offset in current waveforms using digital mimic filtering},
  author={Benmouyal, Gabriel},
  journal={IEEE Transactions on power delivery},
  volume={10},
  number={2},
  pages={621--630},
  year={2002},
  publisher={IEEE}
}

@article{sachdev2007new,
  title={A new algorithm for digital impedance relays},
  author={Sachdev, MS and Baribeau, MA},
  journal={IEEE Transactions on Power Apparatus and Systems},
  number={6},
  pages={2232--2240},
  year={2007},
  publisher={IEEE}
}

@article{jafarian2011weighted,
  title={Weighted least error squares based variable window phasor estimator for distance relaying application},
  author={Jafarian, P and Sanaye-Pasand, M},
  journal={IET generation, transmission \& distribution},
  volume={5},
  number={3},
  pages={298--306},
  year={2011},
  publisher={IET}
}

@article{hwang2018least,
  title={Least error squared phasor estimation with identification of a decaying DC component},
  author={Hwang, Jin Kwon},
  journal={IET Generation, Transmission \& Distribution},
  volume={12},
  number={7},
  pages={1486--1492},
  year={2018},
  publisher={Wiley Online Library}
}

@article{kim2019adaptive,
  title={Adaptive phasor estimation algorithm based on a least squares method},
  author={Kim, Woo-Joong and Nam, Soon-Ryul and Kang, Sang-Hee},
  journal={Energies},
  volume={12},
  number={7},
  pages={1387},
  year={2019},
  publisher={MDPI}
}

@article{sachdev2002recursive,
  title={A recursive least error squares algorithm for power system relaying and measurement applications},
  author={Sachdev, MS and Nagpal, M},
  journal={IEEE Transactions on Power Delivery},
  volume={6},
  number={3},
  pages={1008--1015},
  year={2002},
  publisher={IEEE}
}

@article{da2015phasor,
  title={Phasor estimation in power systems using a neural network with online training for numerical relays purposes},
  author={da Silva, Chrystian Dalla Lana and Cardoso Junior, Ghendy and Mariotto, Lenois and Marchesan, Gustavo},
  journal={IET Science, Measurement \& Technology},
  volume={9},
  number={7},
  pages={836--841},
  year={2015},
  publisher={Wiley Online Library}
}

@article{kim2019study,
  title={A study on deep neural network-based DC offset removal for phase estimation in power systems},
  author={Kim, Sun-Bin and Sok, Vattanak and Kang, Sang-Hee and Lee, Nam-Ho and Nam, Soon-Ryul},
  journal={Energies},
  volume={12},
  number={9},
  pages={1619},
  year={2019},
  publisher={MDPI}
}

@article{sok2022deep,
  title={Deep neural network-based removal of a decaying dc offset in less than one cycle for digital relaying},
  author={Sok, Vattanak and Lee, Sun-Woo and Kang, Sang-Hee and Nam, Soon-Ryul},
  journal={Energies},
  volume={15},
  number={7},
  pages={2644},
  year={2022},
  publisher={MDPI}
}

@INPROCEEDINGS{10591924,
  author={Mohammadi, Sina and Haghighi, Rouzbeh and Hassan, Ali and Bui, Van-Hai and Wang, Mengqi and Su, Wencong},
  booktitle={2024 IEEE 18th International Conference on Control \& Automation (ICCA)}, 
  title={Fast Accurate Phasor Estimation in Less than One Cycle using Neural Networks}, 
  year={2024},
  volume={},
  number={},
  pages={359-363},
  doi={10.1109/ICCA62789.2024.10591924}}

@article{petersen2019deep,
  title={Deep symbolic regression: Recovering mathematical expressions from data via risk-seeking policy gradients},
  author={Petersen, Brenden K and Landajuela, Mikel and Mundhenk, T Nathan and Santiago, Claudio P and Kim, Soo K and Kim, Joanne T},
  journal={arXiv preprint arXiv:1912.04871},
  year={2019}
}

@article{javadi2025review,
  title={A review on symbolic regression in power systems: Methods, applications, and future directions},
  author={Javadi, Amir Bahador and Pong, Philip},
  journal={Renewable and Sustainable Energy Reviews},
  volume={224},
  pages={116075},
  year={2025},
  publisher={Elsevier}
}

@book{rogers2000power,
  title={Power system oscillations},
  author={Rogers, Graham},
  volume={106},
  year={2000},
  publisher={Springer}
}

@article{mohammadi2022decaying,
  title={Decaying DC offset current mitigation in phasor estimation applications: A review},
  author={Mohammadi, Sina and Mahmoudi, Amin and Kahourzade, Solmaz and Yazdani, Amirmehdi and Shafiullah, GM},
  journal={Energies},
  volume={15},
  number={14},
  pages={5260},
  year={2022},
  publisher={MDPI}
}

@article{mohammadi2022novel,
  title={A novel analytical method for DC offset mitigation enhancing DFT phasor estimation},
  author={Mohammadi, Sina and Rezaei, Navid and Mahmoudi, Amin},
  journal={Electric Power Systems Research},
  volume={209},
  pages={108036},
  year={2022},
  publisher={Elsevier BV}
}

@ARTICLE{10315150,
  author={Hollweg, Guilherme Vieira and Bui, Van-Hai and Silva, Felipe Leno Da and Glatt, Ruben and Chaturvedi, Shivam and Su, Wencong},
  journal={IEEE Open Access Journal of Power and Energy}, 
  title={An RMRAC With Deep Symbolic Optimization for DC–AC Converters Under Less-Inertia Power Grids}, 
  year={2023},
  volume={10},
  number={},
  pages={629-642},
  doi={10.1109/OAJPE.2023.3332227}}

@ARTICLE{10215513,
  author={Silva, Felipe Leno da and Glatt, Ruben and Su, Wencong and Bui, Van-Hai and Chang, Fangyuan and Chaturvedi, Shivam and Wang, Mengqi and Murphey, Yi Lu and Huang, Can and Xue, Lingxiao and Zeng, Rong},
  journal={IEEE Journal of Emerging and Selected Topics in Industrial Electronics}, 
  title={AutoTG: Reinforcement Learning-Based Symbolic Optimization for AI-Assisted Power Converter Design}, 
  year={2024},
  volume={5},
  number={2},
  pages={680-689},
  doi={10.1109/JESTIE.2023.3303836}}

\end{document}